\documentclass[aps,reprint,twocolumn,superscriptaddress,citeautoscript,amsmath,amssymb,floatfix,prb]{revtex4-1}
\usepackage{graphicx}
\usepackage{times}
\usepackage{bm}
\usepackage{color}
\usepackage{gensymb}
\usepackage{dcolumn}
\usepackage{xfrac}

\newcolumntype{d}[1]{D{.}{.}{#1}}

\begin{document}
\title{Antiferromagnetic Stacking of Ferromagnetic Layers and Doping Controlled  Phase Competition in Ca$_{1-x}$Sr$_{x}$Co$_{2-y}$As$_{2}$}

\author{Bing~Li}
\affiliation{Ames Laboratory, U.S. DOE, Iowa State University, Ames, Iowa 50011, USA}
\affiliation{Department of Physics and Astronomy, Iowa State University, Ames, Iowa 50011, USA}

\author{Y.~Sizyuk}
\affiliation{Ames Laboratory, U.S. DOE, Iowa State University, Ames, Iowa 50011, USA}
\affiliation{Department of Physics and Astronomy, Iowa State University, Ames, Iowa 50011, USA}

\author{N.~S.~Sangeetha}
\affiliation{Ames Laboratory, U.S. DOE, Iowa State University, Ames, Iowa 50011, USA}

\author{J.~M.~Wilde}
\affiliation{Ames Laboratory, U.S. DOE, Iowa State University, Ames, Iowa 50011, USA}
\affiliation{Department of Physics and Astronomy, Iowa State University, Ames, Iowa 50011, USA}

\author{P.~Das}
\altaffiliation[Present Address:  ]{Institute for Shock Physics, Washington State University, Pullman, WA 99164, USA}
\affiliation{Ames Laboratory, U.S. DOE, Iowa State University, Ames, Iowa 50011, USA}
\affiliation{Department of Physics and Astronomy, Iowa State University, Ames, Iowa 50011, USA}

\author{W.~Tian}
\affiliation{Neutron Scattering Division, Oak Ridge National Laboratory, Oak Ridge, TN 37831, USA}

\author{D.~C.~Johnston}
\affiliation{Ames Laboratory, U.S. DOE, Iowa State University, Ames, Iowa 50011, USA}
\affiliation{Department of Physics and Astronomy, Iowa State University, Ames, Iowa 50011, USA}

\author{A.~I.~Goldman}
\affiliation{Ames Laboratory, U.S. DOE, Iowa State University, Ames, Iowa 50011, USA}
\affiliation{Department of Physics and Astronomy, Iowa State University, Ames, Iowa 50011, USA}

\author{A.~Kreyssig}
\affiliation{Ames Laboratory, U.S. DOE, Iowa State University, Ames, Iowa 50011, USA}
\affiliation{Department of Physics and Astronomy, Iowa State University, Ames, Iowa 50011, USA}

\author{P.~P.~Orth}
\affiliation{Ames Laboratory, U.S. DOE, Iowa State University, Ames, Iowa 50011, USA}
\affiliation{Department of Physics and Astronomy, Iowa State University, Ames, Iowa 50011, USA}

\author{R.~J.~McQueeney}
\affiliation{Ames Laboratory, U.S. DOE, Iowa State University, Ames, Iowa 50011, USA}
\affiliation{Department of Physics and Astronomy, Iowa State University, Ames, Iowa 50011, USA}
\email{rmcqueen@ameslab.gov}

\author{B.~G.~Ueland}
\affiliation{Ames Laboratory, U.S. DOE, Iowa State University, Ames, Iowa 50011, USA}
\affiliation{Department of Physics and Astronomy, Iowa State University, Ames, Iowa 50011, USA}
\email{bgueland@ameslab.gov}

\date{\today}

\begin{abstract}
In search of a quantum phase transition between the two-dimensional ($2$D) ferromagnetism of CaCo$_{2-y}$As$_{2}$ and stripe-type antiferromagnetism in SrCo$_{2}$As$_{2}$, we instead find evidence for $1$D magnetic frustration between magnetic square Co layers. We present neutron diffraction data for Ca$_{1-x}$Sr$_{x}$Co$_{2-y}$As$_{2}$ that reveal a sequence of $x$-dependent magnetic transitions which involve different stacking of $2$D ferromagnetically-aligned layers with different magnetic anisotropy.  We explain the $x$-dependent changes to the magnetic order by utilizing classical analytical calculations of a $1$D Heisenberg model where single-ion magnetic anisotropy and frustration of antiferromagnetic nearest- and next-nearest-layer exchange interactions are all composition dependent.
\end{abstract}

\maketitle

\section{INTRODUCTION}

The $122$-type cobalt pnictides $A$Co$_{2}Pn_{2}$ ($A=$~Ca, Sr, Eu, $Pn=$~As, P) with the tetragonal ThCr$_{2}$Si$_{2}$-type structure (space group $I4/mmm$) \cite{Johnston_2010, Hoffman_1985, Jia_2009, Pandey_2013, Anand_2014, Anand_2014b, Sefat_2009, Imai_2017,Jayasekara_2013} are metals with fascinating properties due to magnetic frustration within their square Co layers \cite{Sapkota_2017} and $Pn$-$Pn$ hybridization-driven magnetoelastic interactions \cite{Hoffman_1985,Reehuis_1998}.  In particular, CaCo$_{1.86}$As$_{2}$ shows evidence of Stoner-enhanced ferromagnetism (FM), and SrCo$_{2}$As$_{2}$  \cite{Pandey_2013, Anand_2014}  harbors itinerant antiferromagnetic (AF) fluctuations \cite{Jayasekara_2013} centered at neutron-momentum transfers $\mathbf{Q}$ corresponding to the stripe-type AF found in various $122$-type Fe-pnictide superconductors \cite{Lynn_2009, Johnston_2010, Paglione_2010, Canfield_2010, Stewart_2011}.  These fluctuations exist despite SrCo$_{2}$As$_{2}$ remaining paramagnetic (PM) down to a temperature of at least $T=0.05$~K \cite{Li_un}.  Here, we present results from neutron diffraction experiments made on the series Ca$_{1-x}$Sr$_{x}$Co$_{2-y}$As$_{2}$ which detail the microscopic changes to the AF order as Ca is replaced  by Sr.

\begin{figure}[]
	\centering
	\includegraphics[width=1.0\linewidth]{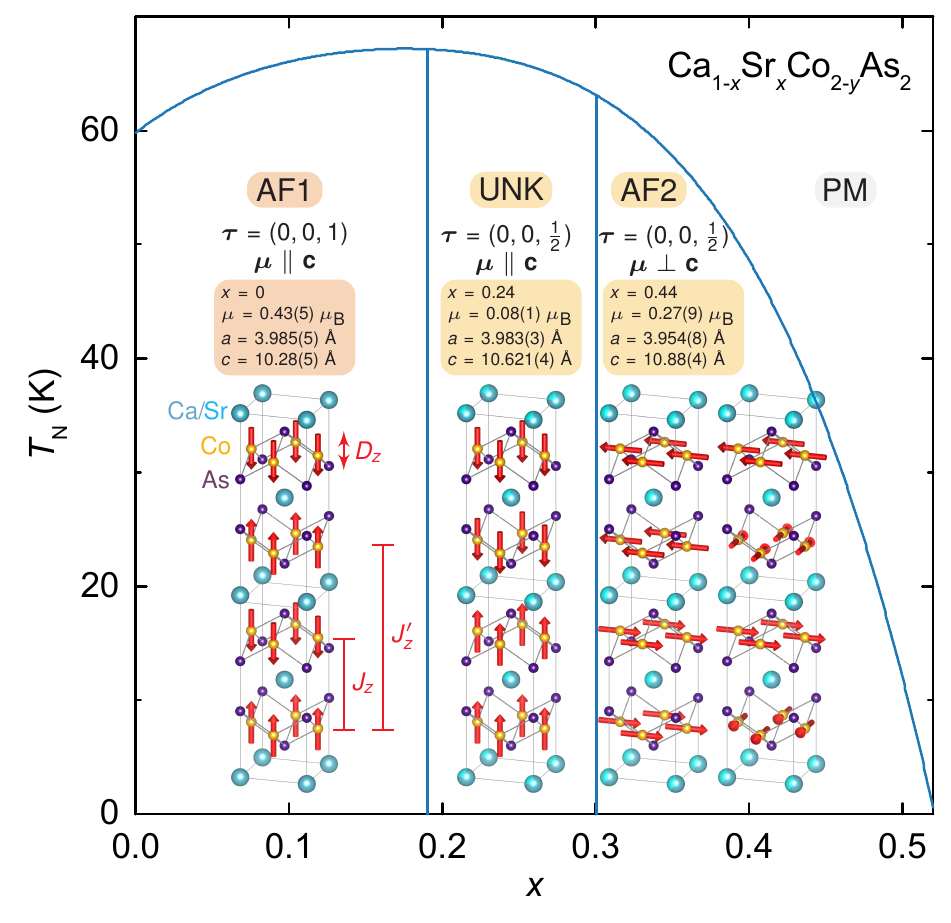}
	\caption{  \label{Fig1} Chemical and magnetic structures of Ca$_{1-x}$Sr$_{x}$Co$_{2-y}$As$_{2}$, and the magnetic phase diagram based on magnetization data from Ref.~[\onlinecite{Sangeetha_2017}]. The AF$1$ phase  has A-type antiferromagnetic (AF) order characterized by  an AF propagation vector of $\bm{\tau}=(0,0,1)$, and consists of ferromagnetic Co layers stacked AF along the crystalline $\mathbf{c}$ axis (${+}{-}{+}{-}$). The ordered magnetic moment $\bm{\mu}$ lies parallel to $\mathbf{c}$. The UNK phase has a  ${+}{+}{-}{-}$ structure with $\bm{\tau}=(0,0,\frac{1}{2})$ and $\bm{\mu}\parallel\mathbf{c}$. The AF$2$ phase has $\bm{\tau}=(0,0,\frac{1}{2})$ but $\bm{\mu}\perp\mathbf{c}$.  Its order is either ${+}{+}{-}{-}$ (left), or a clock-type AF structure (right). The occurrence of magnetic domains prevents us from distinguishing between these structures.  Similarly, an amplitude-modulated spin-density wave cannot be ruled out.  PM stands for paramagnetic and $\mu$ is given per Co atom.  The diagrams were created with \textsc{vesta} [\onlinecite{Momma_2011}].}
\end{figure}

CaCo$_{1.86}$As$_{2} $ has A-type AF order below a N\'{e}el temperature of $T_{\text{N}} = 52$~K, which consists of FM aligned square Co layers stacked AF along the crystalline $\mathbf{c}$ direction (${+}{-}{+}{-}$ structure) and an ordered magnetic moment $\bm{\mu}$ lying parallel to $\mathbf{c}$ \cite{Quirinale_2013, Jayasekara_2017}, as shown in Fig.~\ref{Fig1} \cite{Sangeetha_2017, Momma_2011}.  Despite the A-type order, its spin fluctuation spectrum indicates strong itinerant fluctuations within the extremely magnetically frustrated Co planes and only weak coupling along $\mathbf{c}$ \cite{Sapkota_2017}.  Creating a slight imbalance of the frustrated intralayer interactions of the effectively $2$D FM order may drive the ground state towards stripe-type AF \cite{Sapkota_2017}.  Thus, studying Ca$_{1-x}$Sr$_{x}$Co$_{2-y}$As$_{2}$ provides a possible route to find quantum critical behavior between $2$D FM and stripe-type AF ground states.  

Magnetization data for the Ca$_{1-x}$Sr$_{x}$Co$_{2-y}$As$_{2}$ series reveal multiple magnetic transitions \cite{Ying_2013,Sangeetha_2017} between $x=0$ and $1$, and similar data exist for Ca$_{1-x}$Sr$_{x}$Co$_{2}$P$_{2}$ \cite{Jia_2009}.  Recent reports for the arsenide series find that the $x=0$ A-type phase (AF$1$) transitions to an unknown magnetic phase (UNK) for $0.2 \alt x \alt 0.3$, into an AF phase (AF$2$) with $\bm{\mu}\perp\mathbf{c}$  for $0.3 \alt x\alt 0.5$, and finally into a PM state for $x\agt0.5$.  The effective magnetic anisotropy has different signs in the AF$1$ and AF$2$ phases, with $\bm{\mu}$ oriented parallel or perpendicular to $\mathbf{c}$, respectively.   In the UNK phase, $\bm{\mu}\parallel\mathbf{c}$ but the magnetic anisotropy is essentially zero \cite{Sangeetha_2017}.  The microscopic details of the magnetic order in the UNK and AF$2$ phases were previously unknown.

In this paper, we show that the square Co layers of Ca$_{1-x}$Sr$_{x}$Co$_{2-y}$As$_{2}$ remain FM aligned in the UNK and AF$2$ phases, and that the AF stacking of the layers changes with $x$.  Previous research has shown that materials with such coupled FM aligned planes may be described using a one-dimensional ($1$D) Heisenberg model  \cite{Majlis_2000} in which tuning the interlayer coupling strengths gives rise to a variety of collinear and noncollinear magnetic ground states \cite{Orbach_1958, Yang_1966a, Yang_1966b, Yang_1966c, Baxter_1972}.  We show that this is the case here, and reveal that the evolution of magnetic order in Ca$_{1-x}$Sr$_{x}$Co$_{2-y}$As$_{2}$ can be understood in terms of the above $1$D Heisenberg model with nearest-layer (NL) and next-nearest-layer (NNL) exchange interactions and single-ion magnetic anisotropy.  Using neutron diffraction data and analytical calculations,  we show that the AF ordered phases for $0.2\alt x\alt0.3$ and $0.3\alt x\alt0.5$ \cite{Sangeetha_2017} both have an AF propagation vector of $\bm{\tau}_{\sfrac{1}{2}}\equiv(0,0,\frac{1}{2})$, which requires relatively large NNL exchange over much of the phase diagram.   For $0.2\alt x\alt0.3$ the FM NL exchange is partially frustrated by the AF NNL exchange, which may explain the occurrence of substantial FM correlations \cite{Sangeetha_2017} and a small $\mu$.

\section{EXPERIMENT}
We synthesized plate-like single crystals of Ca$_{1-x}$Sr$_{x}$Co$_{2-y}$As$_{2}$ by solution growth using Sn flux, and confirmed their stoichiometry via energy-dispersive x-ray spectroscopy measurements.  We found no evidence for vacancies of the Co sites in the $x=0.24$ and $0.44$ samples used for the neutron diffraction measurements within an uncertainty of $\approx4.5\%$ (i.e.\ $y=0\pm4.5\%$).  We previously discussed that the presence of vacancies and/or the growth technique used may lead to the different observed values of $T_{\text{N}}$ for $x=0$, ranging from $T_{\text{N}}=52$ to $76$~K \cite{Jayasekara_2017}.  Nevertheless, this level of vacancies does not affect the occurrence of A-type AF order \cite{Jayasekara_2017,Quirinale_2013,Anand_2014,Cheng_2012, Ying_2013, Ying_2012}.  

Neutron diffraction experiments were performed with the HB-$1$A fixed-incident-energy triple-axis spectrometer at the High Flux Isotope Reactor, using a fixed neutron energy of $14.6$~meV.  Effective collimations of $40^{\prime}{-}40^{\prime}{-}40^{\prime}{-}80^{\prime}$ were utilized and pyrolitic graphite  filters were placed before the sample. Single crystals with $x=0.24(3)$ and $0.44(7)$ and masses of $21.9$ and $20.4$~mg, respectively, were measured with their ($H~H~L$) reciprocal-lattice planes coincident with the scattering plane, and cooled down to $T=5$~K using a He closed-cycle refrigerator.  A high-energy x-ray diffraction measurement was made as described in Ref.~[\onlinecite{Jayasekara_2017}] on a $0.6$~mg single-crystal of Ca$_{0.60(2)}$Sr$_{0.40(2)}$Co$_{1.93(3)}$As$_{2}$  at station $6$-ID-D at the Advanced Photon Source to confirm that the sample retained $I4/mmm$ symmetry down to $T=5$~K.  In this report, we express $\mathbf{Q}$ in reciprocal-lattice units (r.l.u.).  

\section{Results}

\begin{figure*}[]
	\centering
	\includegraphics[width=1.0\linewidth]{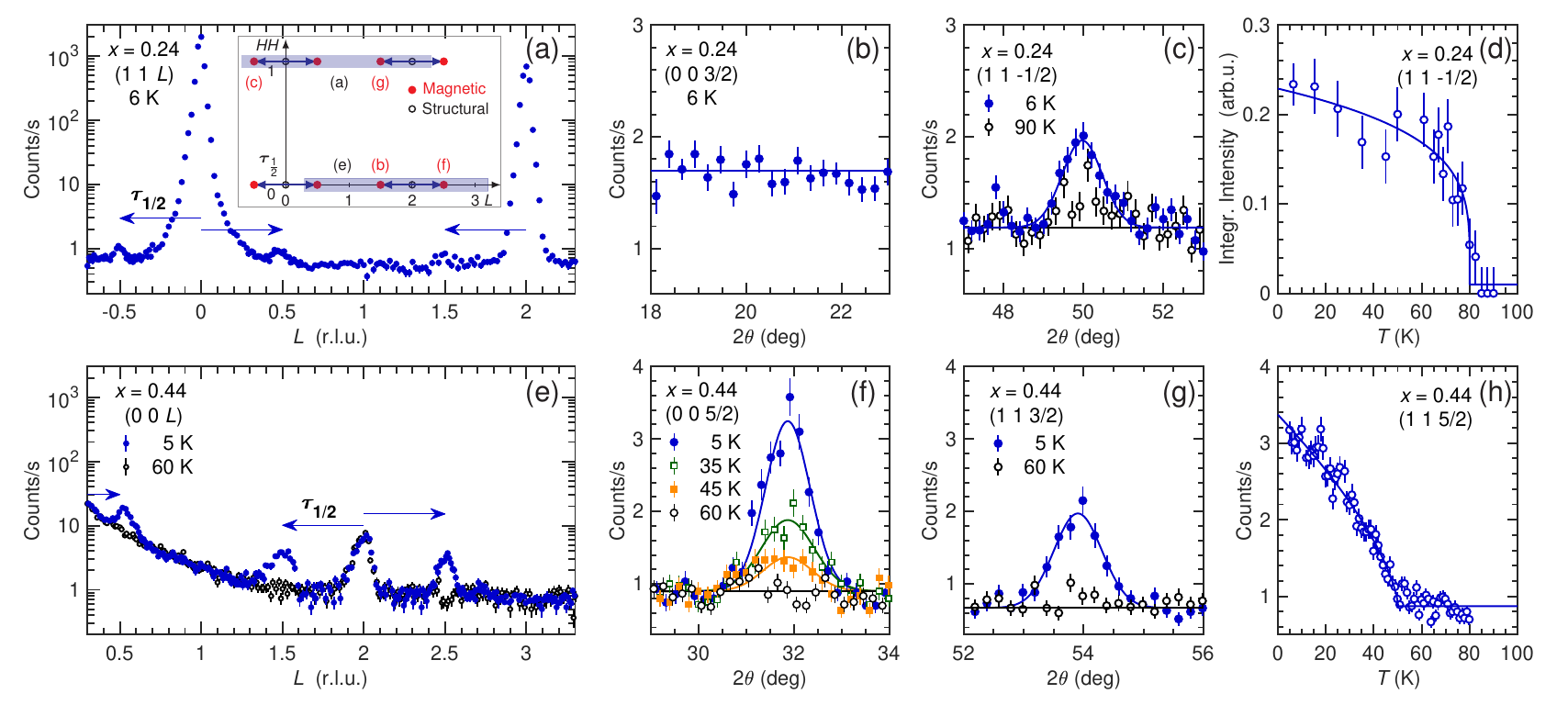}
	\caption{  \label{Fig2} (a-c) Neutron diffraction data for $x=0.24$ from scans along $(1~1~L)$ taken at $T=6$~K (a), and from longitudinal ($\theta$-$2\theta$) scans for $(0~0~\frac{3}{2})$ (b) at $T=6$~K and $(1~1~{-\frac{1}{2}})$ (c) at $T=6$ and $90$~K.  The inset to (a) shows a reciprocal-space map which indicates the scan (grayed areas) or position corresponding to each panel.  (d) Temperature dependence of the integrated intensity of the $x=0.24$ sample's $(1~1~{-\frac{1}{2}})$ magnetic Bragg peak. (e-g) Neutron diffraction data for $x=0.44$ from scans along $(0~0~L)$ taken at $T=5$ and $60$~K (e), and from longitudinal ($\theta$-$2\theta$) scans for $(0~0~\frac{5}{2})$ (f) and $(1~1~{-\frac{1}{2}})$ (g) performed at various temperatures.  (h) Temperature dependence of the intensity of the $x=0.44$ sample's $(1~1~\frac{5}{2})$ magnetic Bragg peak.  Arrows in (a) and (e) point from structural to magnetic Bragg peaks.  Lines in (b), (c), (f), and (g) show fits to a Gaussian lineshape with a constant background, and lines in (d) and (h) are guides to the eye.  }
\end{figure*}

Neutron diffraction data for $x=0.24$ and $0.44$ are shown in Figs.~\ref{Fig2}(a)--\ref{Fig2}(d) and \ref{Fig2}(e)--\ref{Fig2}(h), respectively.  The arrows in Figs.~\ref{Fig2}(a) and \ref{Fig2}(e) point from structural to magnetic Bragg peaks and demonstrate that the AF order is characterized  by $\bm{\tau}_{\sfrac{1}{2}}=(0,0,\frac{1}{2})$ for both compounds.  This propagation vector corresponds to a periodicity along $\mathbf{c}$ of four square Co layers, which differs from the A-type AF order found for $x=0$ with $\bm{\tau_{\text{A}}}=(0,0,1)$ (alternating square Co layers) \cite{Quirinale_2013}, and from the stripe-type AF order found in many $122$-type Fe-pnictide superconductors and the stripe-type spin fluctuations in SrCo$_{2}$As$_{2}$ with $\bm{\tau_{\text{st}}}=(\frac{1}{2},\frac{1}{2},1)$ \cite{Jayasekara_2013}.   The widths of the magnetic and structural Bragg peaks are similar, which attests to the presence of long-range AF order.  We find no evidence for magnetic Bragg peaks in data measured at reciprocal-lattice positions corresponding to $\bm{\tau_{\text{A}}}$ and $\bm{\tau_{\text{st}}}$.

Figures~\ref{Fig2}(b) and \ref{Fig2}(e), respectively, illustrate that magnetic Bragg peaks are absent at $\mathbf{Q}=(0,0,\frac{L}{2})$, $L=$ odd integer, positions for $x=0.24$, but that they occur for $x=0.44$.  Since neutron diffraction is sensitive to the component of $\bm{\mu}\perp\mathbf{Q}$, these data indicate that $\bm{\mu}\parallel\mathbf{c}$ for $x=0.24$, whereas $\bm{\mu}$ has a component in the $\mathbf{ab}$ plane for $x=0.44$.  These results agree with the conclusions from magnetization data that $\bm{\mu}\parallel\mathbf{c}$ for $x=0.24$ and $\bm{\mu}\perp\mathbf{c}$ for $x=0.44$ \cite{Sangeetha_2017}.

Figures~\ref{Fig2}(c), \ref{Fig2}(f), and \ref{Fig2}(g) show detailed views of select magnetic Bragg peaks at various temperatures.  The peaks are quite weak, which implies small ordered moments.  Using the magnetic structures shown in Fig.~\ref{Fig1} for the UNK and AF$2$ phases, we find  $\mu=0.08(1)~\mu_{\text{B}}/$Co for $x=0.24$ at $T=6$~K and $0.27(9)~\mu_{\text{B}}/$Co for $x=0.44$ at $5$~K.  Three magnetic Bragg peaks were used to determine $\mu$ for $x=0.24$ and six were used for $x=0.44$.  More details concerning the magnetic structures are given below.

The temperature dependence of the magnetic order parameter for $x=0.24$ and $0.44$ is presented in Figs.~\ref{Fig2}(d) and \ref{Fig2}(h), respectively.  Upon cooling, a magnetic diffraction signal first appears at $T_{N}\approx80$ and $50$~K for $x=0.24$ and $0.44$, respectively, which are above the values of $T_{\text{N}}\approx67$ and $37$~K expected from magnetization data \cite{Sangeetha_2017}.  This may be due to differences in the level of vacancies between the samples used for the neutron diffraction experiments and those used for magnetization \cite{Sangeetha_2017}.  In particular, the level of Co vacancies for the neutron diffraction samples is $y=0.00(4)$, whereas the magnetization samples with $x=0.25$ and $0.45$ have $y=0.10(5)$ and $0.08(2)$, respectively. Samples with Co vacancies seem to have lower values for $T_{\text{N}}$, with $T_{\text{N}}=52(1)$~K for $y=0.14$ \cite{Quirinale_2013} and $T_{\text{N}}=76$~K for $y=0$ \cite{Cheng_2012}.  On the other hand, it has been suggested that vacancies alone may not explain the differing values of $T_{\text{N}}$ from different reports, and that different growth conditions may also be responsible \cite{Zhang_2015}.

\section{Discussion}
We capture the observed magnetic ordering behaviors using the classical Heisenberg Hamiltonian:
\begin{align}
\mathcal{H}&=\mathcal{H}_{\text{in-plane}} + J_{z}\sum_{\mathbf{R}}{\mathbf S}_{\mathbf{R}}\cdot{\mathbf S}_{\mathbf{R}+\mathbf{d}}+J_{z}^{\prime}\sum_{\mathbf{R}}{\mathbf S}_{\mathbf{R}}\cdot{\mathbf S}_{\mathbf{R}+2\mathbf{d}}\nonumber\\
&-D_{z}\sum_{\mathbf{R}}\left( S^z_{\mathbf{R}}\right)^2-D_{xy}\sum_{\mathbf{R}}\left[\left( S^x_{\mathbf{R}}\right)^4+\left( S^y_{\mathbf{R}}\right)^4\right] \,.
\label{eq:Hamiltonian}
\end{align}
Here, $\mathcal{H}_{\text{in-plane}}$ contains competing FM and AF interactions between Heisenberg spins within a square layer, $J_{z}$ ($J_{z}^{\prime}$) is the effective NL (NNL) magnetic exchange interaction along $\mathbf{c}$, $D_{z}$ ($D_{xy}$) is the single-ion magnetic anisotropy along $\mathbf{c}$ (within the $\mathbf{ab}$ plane), and $\mathbf{d}=d\mathbf{\hat{c}}$ where $d$ is the distance between neighboring Co layers.  We regard each FM-aligned Co layer as a single localized Heisenberg spin ${\mathbf S}_{\mathbf{R}}$ at position $\mathbf{R}$ and consider the layers' relative orientations along $\mathbf{c}$ in terms of a $1$D model.  For helical AFs, this is a common model denoted as the $J_{0}$-$J_{1}$-$J_{2}$ model \cite{Johnston_2012, Johnston_2015}.

\begin{figure}[]
	\centering
	\includegraphics[width=1.0\linewidth]{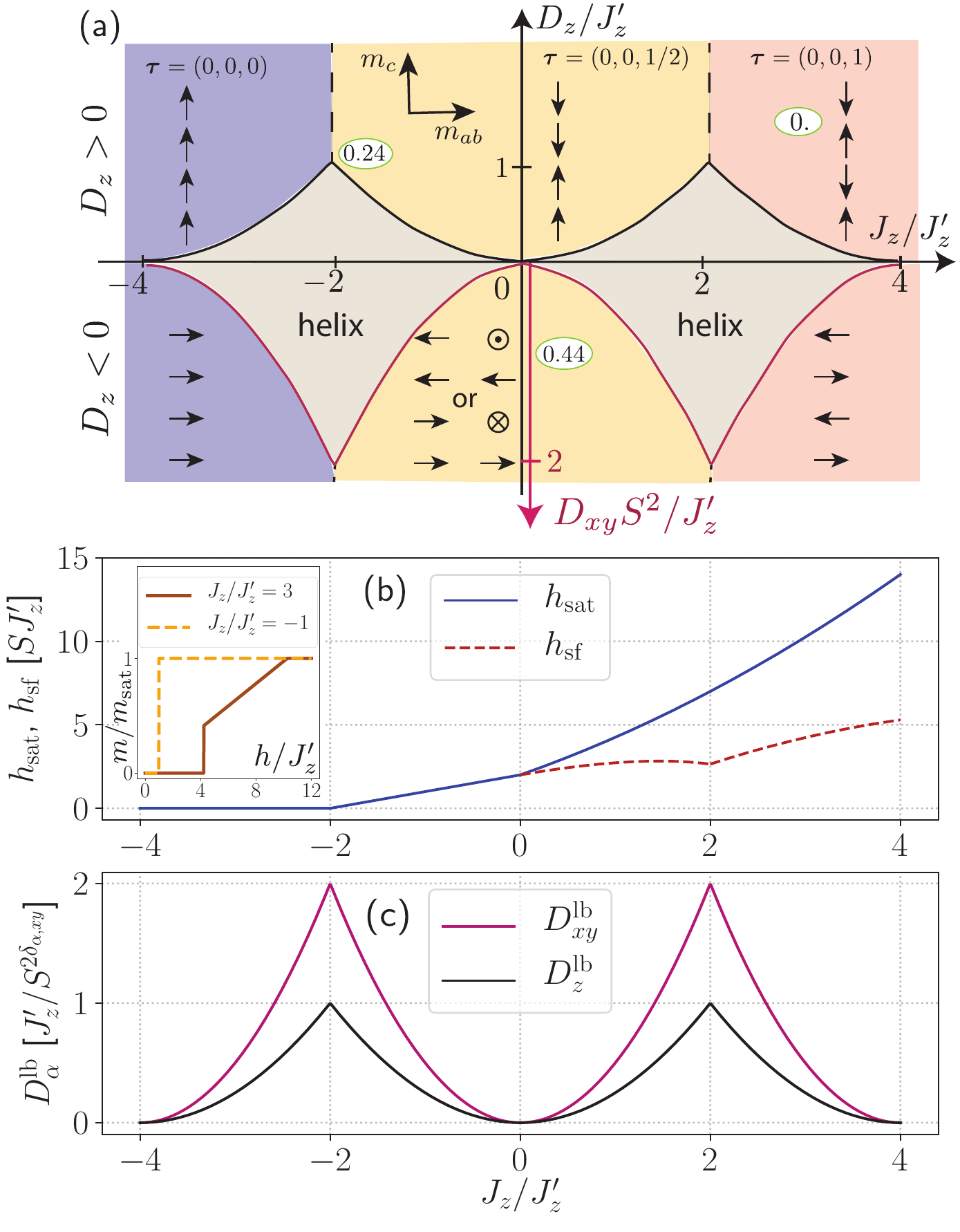}
	\caption{  \label{Fig3} (a) Zero-temperature magnetic phase diagram for varying nearest-layer exchange $J_{z}$ and single-ion magnetic anisotropy $D_{z}$ along $\mathbf{c}$, and fixed antiferromagnetic (AF) next-nearest-layer exchange $J^{\prime}_{z}$ ($J^{\prime}_{z} >0$).  The lower bound of  $D_{z}$, $D_{z}^{\text{lb}}$, and anisotropy within the $\mathbf{ab}$ plane, $D^{\text{lb}}_{xy}$, necessary to suppress helical-AF order are indicated by black and magenta lines, respectively. Arrows illustrate the magnetic order, and numbers in ovals show estimated locations for three values of $x$. (b) Calculated spin-flop ($h_{\text{sf}}$) and saturation ($h_{\text{sat}}$) magnetic fields along $\mathbf{c}$ versus $J_{z}/J^{\prime}_{z}$ for $D_{z}/J^{\prime}_{z} = 1$ and $D_{xy}=0$.  The fields correspond to using $J^{\prime}_{z} = 3 J_{z}$ for $x=0$. The inset shows the calculated magnetization $m$ versus magnetic field $h$ curves for $J_{z}/J_{z}^{\prime} = -1$~(yellow) and $3$~(red). (c) $D^{\text{lb}}_{z}$ and $D^{\text{lb}}_{xy}$ in units of $J_{z}^{\prime}$ and  $J_{z}^{\prime}/S^{2}$, respectively, versus $J_{z}/J^{\prime}_{z}$.}
\end{figure}

We analytically calculate the classical ground-state energies in units of $J_{z}^{\prime}$, which we assume to be AF ($J_{z}^{\prime}>0$), and find the phase diagram given in Fig.~\ref{Fig3}(a).  More details of our calculations are given in Appendix~\ref{AppA}.  In the absence of anisotropy, the ground state is either an  A-type AF, a single-${\mathbf{Q}}$ helix with a turn angle of $\phi = \cos^{-1}[-J_{z}/(4J^{\prime}_{z})]$, or FM, with phase boundaries at $J_{z}/J_{z}^{\prime}=4$ and $-4$, respectively \cite{Johnston_2012,Johnston_2015, Johnston_2017}.  AF order with a propagation vector of $\bm{\tau}_{\sfrac{1}{2}}$ occurs only at $J_{z}/J_{z}^{\prime}=0$.

On the other hand, both neutron diffraction and magnetization data require that $D_{z}>0$ for  $x \alt 0.3$ and $D_{z}<0$ for $0.3\alt x \alt 0.5$, and it turns out that magnetic anisotropy suppresses helical AF order in favor of regions with either $\bm{\tau}_{\sfrac{1}{2}}$-type AF, A-type AF, or FM order.  In the case of $D_{z}>0$, $D_{z}$ must be greater than a lower bound of $D_{z}^{\text{lb}}$ to suppress helical AF order.  For $D_{z}<0$, anisotropy that picks a specific direction within the $\mathbf{ab}$ plane, $D_{xy}$, must be included and must be greater than a lower bound of $D_{xy}^{\text{lb}}$ to suppress helical AF order.  We determined $D_{z}^{\text{lb}}$ and  $D_{xy}^{\text{lb}}$ by comparing the energy of a helical AF state at finite $D_{z}>0$ or $D_{xy}$ (with $D_{z}<0$) to the state found at large $D_{z}$ or $D_{xy}$ (with $D_{z}<0$).  The computed boundaries are plotted in Fig.~\ref{Fig3}(c) and are included in Fig.~\ref{Fig3}(a), where the helix region corresponds to coplanar helical AF order.  

The top part of Fig.~\ref{Fig3}(a) shows that for $D_{z}>D_{z}^{\text{lb}}$, the value of $\lvert J_{z} \rvert/J_{z}^{\prime}$ and the sign of $J_{z}$ determine the stacking of the FM layers.  The ground state is ${+}{+}{-}{-}$ ($\bm{\tau}_{\sfrac{1}{2}}$) for $\lvert J_{z} \rvert/J_{z}^{\prime} < 2$ and either FM (${+}{+}{+}{+}$) or A-type AF (${+}{-}{+}{-}$) for $\lvert J_{z}\rvert/J_{z}^{\prime} >2$.  Note that half of the NL interactions are frustrated for $\lvert J_{z} \rvert/J_{z}^{\prime}<2$ whereas the NNL interactions are frustrated for $\lvert J_{z} \rvert/J_{z}^{\prime}>2$.  In other words, $J_{z}^{\prime}$ dominates $J_{z}$ for $\lvert J_{z} \rvert/J_{z}^{\prime} < 2$ and $J_{z}$ dominates $J_{z}^{\prime}$ for $\lvert J_{z} \rvert/J_{z}^{\prime} > 2$.  Figure~\ref{Fig3}(a) shows that the phase diagram looks quite similar for $D_{z}<0$ and $D_{z}>0$. 

Our theory predicts that $D_{z}>D_{z}^{\text{lb}}$ for $x=0.24$ and that the corresponding AF structure is ${+}{+}{-}{-}$ with $\bm{\mu}\parallel \mathbf{c}$, as shown in Fig.~\ref{Fig1} for the UNK phase.  For  $x=0.44$, which has $D_{z}<0$, we predict $D_{xy} > D^{\text{lb}}_{xy}$ and that the AF order is either the ${+}{+}{-}{-}$ or the $4$-state clock structure shown for the AF$2$ phase in Fig.~\ref{Fig1}.  Both of these  magnetic structures correspond to $\bm{\tau}_{\sfrac{1}{2}}$ and produce similar neutron diffraction patterns due to the presence of magnetic domains. We cannot differentiate between them using our data.  Similarly, we cannot rule out an amplitude-modulated spin-density wave for either $x=0.24$ or $0.44$.  The absence of evidence for a distortion away from tetragonal symmetry in our high-energy x-ray diffraction data for $x=0.40$ may favor the $4$-state clock structure for AF$2$.

Figure~\ref{Fig3}(a) also illustrates that close to $J_{z}/J^{\prime}_{z} = \pm 2$ the degree of anisotropy needed to suppress helical AF order becomes quite significant.  In particular, for $D_{z}<0$ and a weak dependence of $D_{xy}$ on $x$,  we predict distorted helix states to emerge  when $J_{z}/J_{z}^{\prime}$ falls outside the window where $\bm{\tau}_{\sfrac{1}{2}}$-type order is stabilized. These distorted helix states are multi-${\mathbf Q}$ states, and have a turn angle which varies along $\mathbf{c}$ as the spins are canted towards $\mathbf{a}$ and $\mathbf{b}$ by $D_{xy}$.  The precise form of the distorted helix order, which is not observed for $x=0.44$, may be determined numerically, as done for helical AFs in Refs.~[\onlinecite{Johnston_2017}] and [\onlinecite{Johnston_arxiv_2018}], or by classical Monte-Carlo simulations.

We further test our model by analytically determining the spin-flop and saturation magnetic fields, $h_{\text{sf}}$ and $h_{\text{sat}}$, respectively, for the A-type and $\bm{\tau_{\sfrac{1}{2}}}$-type ground states with $D_{z}>D_{z}^{\text{lb}}$, and compare its predictions to magnetization $M$ versus magnetic field $H$ data \cite{Sangeetha_2017}.  We assume that the $\mathbf{ab}$ component of the flopped spins have helical AF order similar to that for $D_{z}<0$ and negligible $D_{xy}$, and plot the results in Fig.~\ref{Fig3}(b).  In our calculations, $h=g\mu_{\text{B}}H$ and we set $\mu=g\mu_{\text{B}}S=1$, where $g$ is the spectroscopic splitting factor.

We find that our model predicts the observed spin flop in the A-type AF phase \cite{Sangeetha_2017} for $2 <J_{z}/J_{z}^{\prime} < 4$ with
\begin{equation}
h_{\text{sf}}=\frac{SJ_{z}^{\prime}}{4}\sqrt{8d_{z}- (j_{z}-4)^2} \sqrt{-8d_{z}+ (j_{z}+4)^2}\,,
\end{equation}
where $j_{z} = J_{z}/J_{z}^{\prime}$ and $d_{z} = D_{z}/J_{z}^{\prime}$.  A spin flop occurs only if $j_{z} > \sqrt{8\left(d_{z}-2\right)}$; otherwise the compound directly saturates with increasing $H$.  For $-2 < j_{z} < 2$, the spin-flop field is
\begin{equation}
h_{\text{sf}} =\frac{S J_{z}^{\prime}}{4}\sqrt{8d_{z}-j_{z}^2} \sqrt{-8d_{z}+(j_{z}+4)^2} \,, 
\end{equation}
and a spin flop occurs only for $j_{z} > -2+2\sqrt{2 d_{z}-1}$.

For the saturation fields, we find ${h_{\text{sat}}=\frac{SJ_{z}^{\prime}}{4}\left[-8d_{z}+(j_{z}+4)^2\right]}$ for $-2 <J_{z}/J_{z}^{\prime} < 4$.  If the system directly saturates without a spin flop, we find $h_{\text{sat}}=2SJ_{z}$ for $2 <J_{z}/J_{z}^{\prime} < 4$ and $h_{\text{sat}}=S(J_{z}+2J_{z}^{\prime})$ for $-2 <J_{z}/J_{z}^{\prime} < 2$.  

Using the expressions for $2 <J_{z}/J_{z}^{\prime} < 4$, the experimental $M(H)$ data for $x=0$ \cite{Sangeetha_2017} and the experimentally determined value of $\mu=0.43~\mu_{\text{B}}/\text{Co}$ \cite{Jayasekara_2017}, we estimate that $J_{z} \approx 0.24$~meV and $D_{z} \approx 0.08$~meV for $x=0$ and place it on the phase diagram in Fig.~\ref{Fig3}(a) by arbitrarily assuming that $J_{z}/J_{z}^{\prime}=3$.  We place the $x=0.24$ compound in the $\bm{\tau}_{\sfrac{1}{2}}$-type AF ordered region of Fig.~\ref{Fig3}(a) corresponding to FM $J_{z}$ and $D_{z}>D_{z}^{\text{lb}}$ based on the observed AF propagation vector for $x=0.24$ and the fact that magnetization data find evidence for strong FM correlations along $\mathbf{c}$ coexisting with the AF order \cite{Sangeetha_2017}.  In this region, $J_{z}$ is partially frustrated, and may cause the strong FM correlations and a value of $\mu$ much lower than that found for either $x=0$ or $0.44$.  Magnetization data for $x=0.44$ do not show evidence for strong FM correlations along $\mathbf{c}$ \cite{Sangeetha_2017}.  Hence, we place it on the positive side of the $J_{z}/J^{\prime}_{z}$ axis in Fig.~\ref{Fig3}(a).

For Ca$_{1-x}$Sr$_{x}$Fe$_{2}$As$_{2}$, stripe-type AF order persists across the series with an almost constant $\bm{\mu}$ despite $T_{\text{N}}$ increasing by $\approx48\%$ between $x=0$ and $0.3$ \cite{Kirshenbaum_2012}.   The change in $T_{\text{N}}$ is tied to changes in the chemical unit cell size and structure.  This behavior is distinct from our observations for Ca$_{1-x}$Sr$_{x}$Co$_{2-y}$As$_{2}$.  Nevertheless, the  crossover from a collapsed-tetragonal to tetragonal phase in Ca$_{1-x}$Sr$_{x}$Co$_{2-y}$As$_{2}$ and the associated large increase in $c$ and changes to other unit-cell parameters with increasing $x$ \cite{Sangeetha_2017} likely play a role in the variation of $J_{z}$ and $D_{z}$ with composition.

A recent report on electronic-band-structure calculations for $A$Co$_{2}$As$_{2}$, $A=$~Ca, Sr, Ba, find that the overall electronic structures are similar for all three compounds, with only the proximity to a van Hove singularity of a flat band associated with the Co $e_{g}$ orbitals responsible for FM fluctuations within the $\mathbf{ab}$ plane differing between them \cite{Mao_2018}.  The flat band lies just below the van Hove singularity for $A=$~Ca, causing an enhancement of the dynamical susceptibility which may lead to the observed A-type AF order.  The flat band lies further away from the van Hove singularity for $A=$~Sr and Ba, which have not been observed to magnetically order.   The closer proximity of the flat band to the van Hove singularity for $A=$~Ca is tied to a larger Co-Co bond length and lower As height above a Co layer.   It would be interesting to observe how the enhancement in the dynamical susceptibility seen for $A=$~Ca changes as Ca is systematically replaced by Sr.

\section{Conclusion}
Our results highlight the manifestation of highly-tunable and analytically-determinable magnetic ground states in Ca$_{1-x}$Sr$_{x}$Co$_{2-y}$As$_{2}$ in the presence of frustrated NL or NNL exchange between FM-aligned square Co layers and magnetic anisotropy.  More generally, we have found that the cobalt-arsenide system manifests strong magnetic frustration both within its square layers and between them.  The origins of frustration within the layers likely trace back to flat electronic bands associated with Stoner-like ferromagnetism \cite{Mao_2018}, whereas here we highlight a different kind of frustration: frustration between FM-aligned layers.  Future band structure calculations and inelastic neutron scattering experiments can provide detailed information on the magnetic state of the layers themselves, and determine whether or not the itinerant FM fluctuations present for $x=0$ persist into the UNK and AF$2$ phases, and if the stripe-type fluctuations found in SrCo$_{2}$As$_{2}$ \cite{Jayasekara_2013} exist in Ca$_{1-x}$Sr$_{x}$Co$_{2-y}$As$_{2}$.  Such work should also result in a better understanding of the microscopic origin of the compositional changes to our Heisenberg model's parameters, as well as the limits of our $1$D local-moment model.

\begin{acknowledgments}
We are grateful for assistance from D.~Robinson with performing the x-ray experiments, and helpful conversations with T. W. Heitmann, D.~H.~Ryan, L.~Ke, and D.~Vaknin.  Work at the Ames Laboratory was supported by the U.~S.~Department of Energy (DOE), Basic Energy Sciences, Division of Materials Sciences \& Engineering, under Contract No.~DE-AC$02$-$07$CH$11358$. A portion of this research used resources at the High Flux Isotope Reactor, a U.~S.~DOE Office of Science User Facility operated by the Oak Ridge National Laboratory.  This research used resources of the Advanced Photon Source, a U.~S.~DOE Office of Science User Facility operated for the U.~S.~DOE Office of Science by Argonne National Laboratory under Contract No.~DE-AC$02$-$06$CH$11357$. Y.~S. and P.~P.~O. acknowledge support from Iowa State University Startup Funds.
\end{acknowledgments}

\appendix
\section{Analytical Calculations}\label{AppA}
\subsection{Introduction}
We model the  stacking along the tetragonal $\mathbf{c}$ axis of the ferromagnetically (FM) aligned square Co layers in Ca$_{1-x}$Sr$_{x}$Co$_{2-y}$As$_{2}$ using the classical local-moment Heisenberg spin Hamiltonian:
\begin{align}
\mathcal{H}&=\mathcal{H}_{\text{in-plane}} + J_{z}\sum_{\mathbf{R}}{\mathbf S}_{\mathbf{R}}\cdot{\mathbf S}_{\mathbf{R}+\mathbf{d}}+J_{z}^{\prime}\sum_{\mathbf{R}}{\mathbf S}_{\mathbf{R}}\cdot{\mathbf S}_{\mathbf{R}+2\mathbf{d}}\nonumber\\
&-D_{z}\sum_{\mathbf{R}}\left( S^z_{\mathbf{R}}\right)^2-D_{xy}\sum_{\mathbf{R}}\left[\left( S^x_{\mathbf{R}}\right)^4+\left( S^y_{\mathbf{R}}\right)^4\right] \,,
\label{eq:HamiltonianApp}
\end{align}
and analytically calculate the classical ground-state energies in units of $J_{z}^{\prime}$, which we assume to be antiferromagnetic (AF) ($J_{z}^{\prime}>0$).   We regard each FM-aligned Co layer as a single Heisenberg spin ${\mathbf S}_{\mathbf{R}}$ at position $\mathbf{R}$,  and consider the layers' relative orientations along $\mathbf{c}$ in terms of a one-dimensional ($1$D) model.  For helical AF, this is a common model denoted as the $J_{0}$-$J_{1}$-$J_{2}$ model \cite{Johnston_2012, Johnston_2015}.  In Eq.~(\ref{eq:HamiltonianApp}), $\mathcal{H}_{\text{in-plane}}$ contains competing FM and AF interactions between Heisenberg spins within a square layer, $J_{z}$ ($J_{z}^{\prime}$) is the effective nearest-layer (NL) [next-nearest-layer (NNL)] exchange along $\mathbf{c}$, $D_{z}$ ($D_{xy}$) is the single-ion magnetic anisotropy along $\mathbf{c}$ (within the $\mathbf{ab}$ plane), and $\mathbf{d}=d\mathbf{\hat{c}}$, where $d$ is the distance between neighboring Co layers.  

The following subsections give details of the calculations for the spin-flop $h_{\text{sf}}$ and saturation $h_{\text{sat}}$ fields in different regions of the phase diagram shown in Fig.~3(a) for a magnetic field $\mathbf{H}$ applied along $\mathbf{c}$.  Details of the calculations used to estimate the lower bounds of $D_{z}$ and $D_{xy}$ necessary to suppress a helix state are also presented. 

To determine $h_{\text{sf}}$ and $h_{\text{sat}}$,  the standard Zeeman-interaction term is added to Eq.~(\ref{eq:HamiltonianApp}):
\begin{equation}
\mathcal{H}_{Z}=-\mu\cdot H= h_z\sum_{\mathbf{R}} S_{\mathbf{R}}^z\,.
\label{eq:Zeeman}
\end{equation}
For brevity, the spectroscopic-splitting factor $g$ and Bohr magneton $\mu_\text{B}$ are absorbed into the field definition $h_z=g\mu_B H_z$, where $H_z$ is the magnetic field applied along $\mathbf{c}$. These constants are restored at the end for numerical estimates of $J_{z}$ and $D_{z}$ that incorporate experimental results.

\subsection{$\bm{J_z/J_z^{\prime}>4}$, $\bm{D_z \geq 0}$, and $\bm{h_z>0}$}
\begin{figure}[]
	\centering
	\includegraphics[width=.4\linewidth]{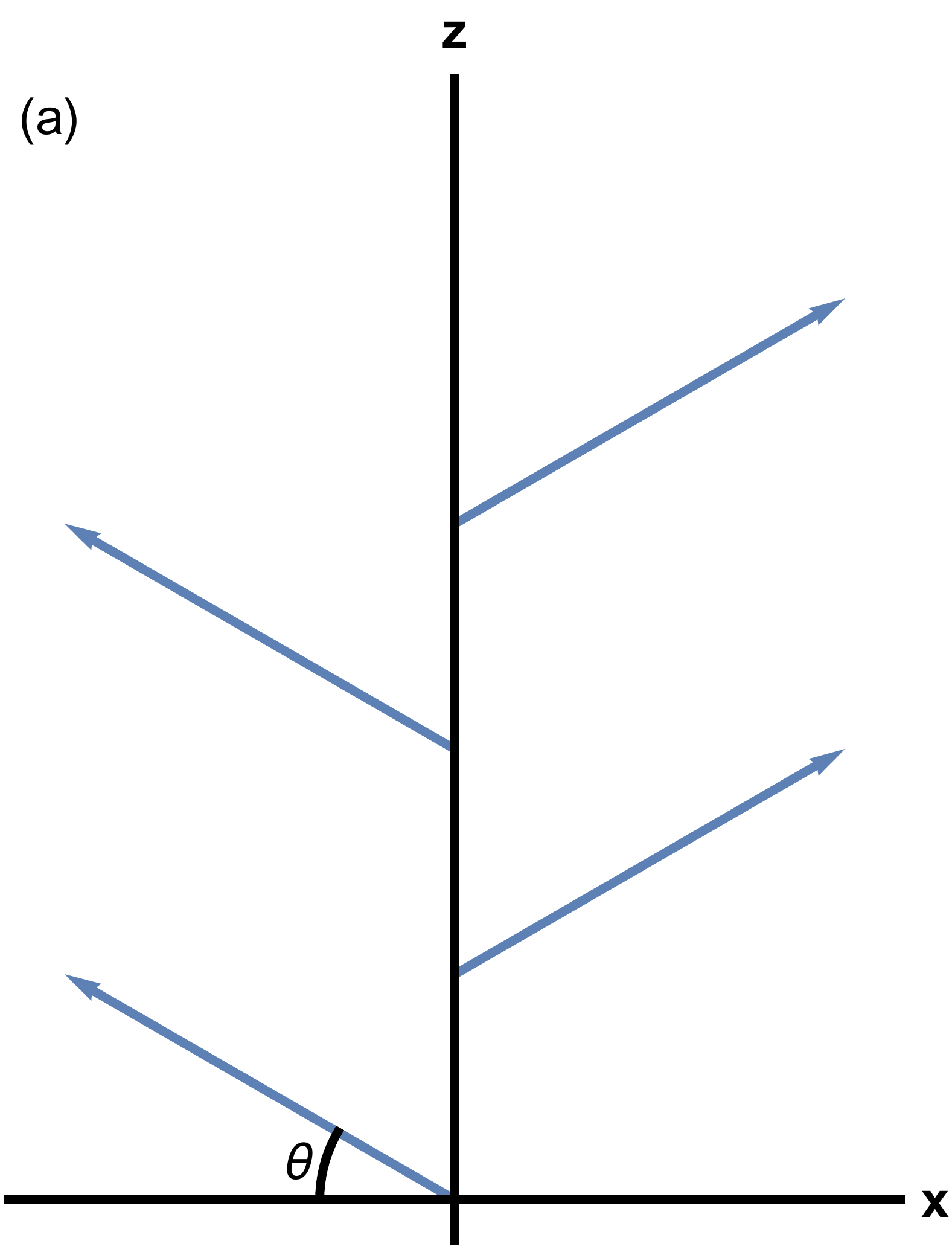}
	\includegraphics[width=.55\linewidth]{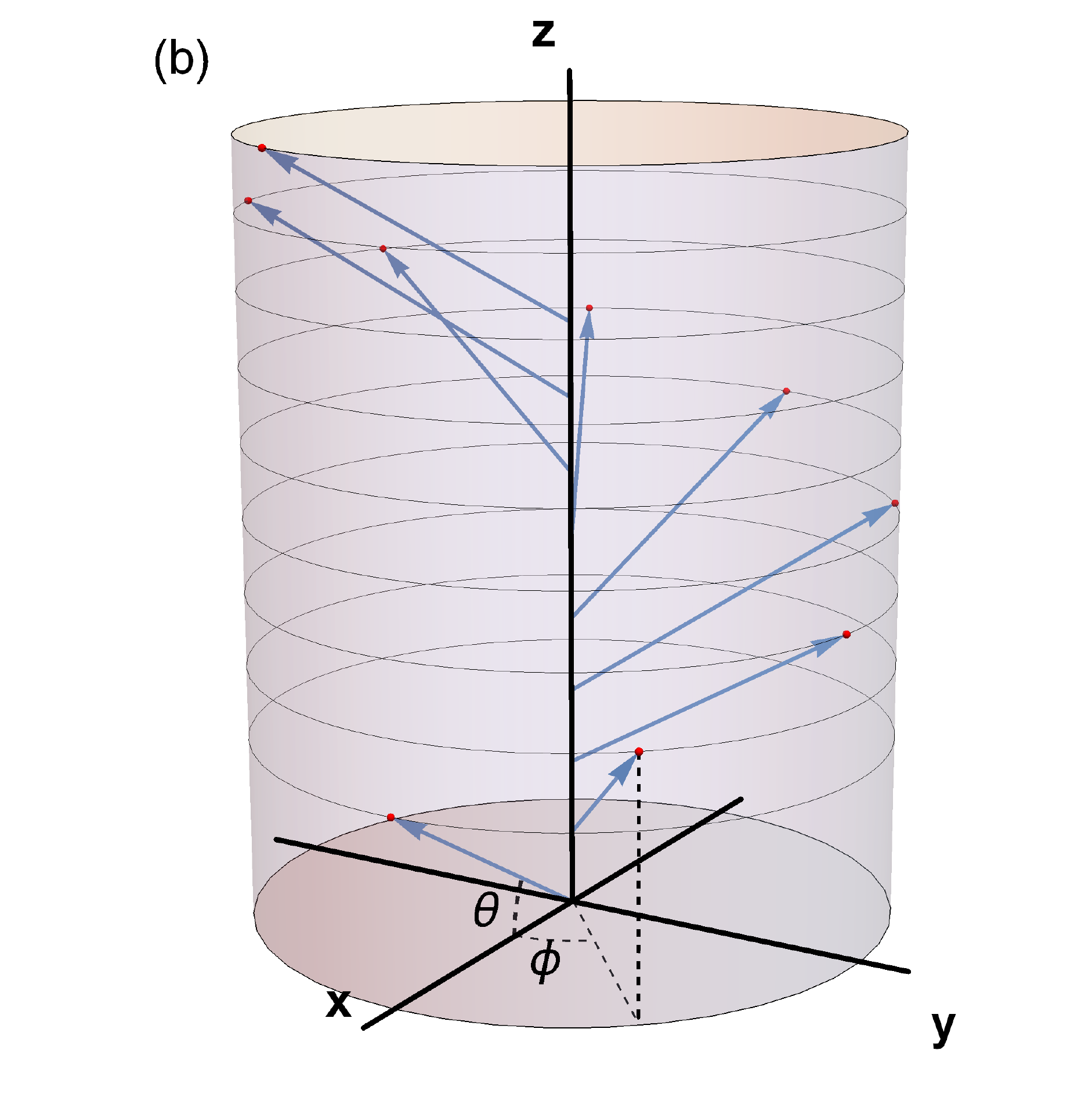}
	\caption{\label{FigS1} Variational ground states of the Hamiltonian in Eq.~(\ref{eq:HamiltonianApp}) with a magnetic field applied along the crystalline $\mathbf{c}$ axis [see Eq.~(\ref{eq:Zeeman})] for (a) $J_z/J_z^{\prime}>4$ and (b) $-4<J_z/J_z^{\prime}<4$. $\theta$ is the canting angle, and $\phi$ is the turning angle of the helix. }
\end{figure}

The ground state in the regime $J_z/J^{\prime}_z > 4$ and $D_z \geq 0$ is A-type AF order with moments laying along $\mathbf{c}$, as shown for the AF$1$ phase in Fig.~1.  The AF propagation vector $\bm{\tau}$ is $\bm{\tau}_{\text{A}}=(0,0,1)$.  We consider the following variational state [see Fig.~\ref{FigS1}(a)]:
\begin{equation}
{\mathbf S}=(\cos\theta\cos[{\mathbf Q}\cdot{\mathbf R}],0,\sin\theta),
\end{equation}
and set $\mathbf{Q}$ equal to $\bm{\tau}_{\text{A}}$. Note that $\theta$ is measured from the $\mathbf{ab}$ plane rather than $\mathbf{c}$.  The energy of this state is
\begin{equation}
E_1=NS^2[-\cos(2\theta)J_z-\sin^2\theta D_z+J_z^{\prime}]-NSh_z\sin\theta \,. \label{eq:E1}
\end{equation}
Minimizing Eq.~(\ref{eq:E1}) with respect to $\theta$  gives:
\begin{equation}
J_z\sin2\theta-D_z\sin\theta\cos\theta=\frac{h_z}{2S}\cos\theta \,,
\end{equation}
which has two solutions:
\begin{align}
\theta_1&=\pi/2 \intertext{and}
\theta_2&=\sin^{-1}{\bigg[\frac{h_z}{2S(2J_z-D_z)}\bigg]}\,.
\end{align}
The first solution is the saturated state, where the spins are fully polarized along the field direction.  The second solution is a canted state, in which the spins are canted away from $\mathbf{c}$.   $h_{\text{sf}}$ is found by setting the energy of the canted state equal to the energy of the $h_z=0$ state. This yields 
\begin{equation}
h_{\text{sf}}=\frac{S}{\sin\theta_2}[J_z(1-\cos2\theta_2)+D_z(1-\sin^2\theta_2)]\,. \label{eq:hsfA}
\end{equation}
Solving Eq.~(\ref{eq:hsfA}) gives
\begin{equation}
h_{\text{sf}}=2S\sqrt{D_z(2J_z-D_z)} \,.\label{eq:hsfA_2}
\end{equation}
Note that $h_{\text{sf}}$ vanishes as $D_z \rightarrow 0$.

The canted state exists only if $h_{\text{sf}}<2S(2J_z-D_z)$.  It must occur before the saturated state occurs, otherwise the spins would directly saturate. We can determine when this happens by setting the energy of the saturated state equal to the $h_z=0$ state, which gives:
\begin{equation}
h_{\text{sat}}=2SJ_z\,.\label{eq:hsatA}
\end{equation}
Comparing Eqs.~(\ref{eq:hsfA_2}) and (\ref{eq:hsatA}), we see that the canted state exists for $J_z>D_z$. To summarize, for the $J_z/J_z^{\prime}>4$ and $D_z \geq 0$ region of the phase diagram we have:
\begin{align}
&h_{\text{sat}}=2SJ_z\,\intertext{for  $J_z<D_z$, and}
&h_{\text{sf}}=2S\sqrt{D_z(2J_z-D_z)} \,,\\
&h_{\text{sat}}=2S(2J_z-D_z)\,
\end{align}
for $J_z>D_z$.

\subsection{ $\bm{-4<J_z/J_z^{\prime}<4}$, $\bm{D_z \geq 0}$, and $\bm{h_z>0}$}
Let us consider the regime with $-4<J_z/J_z^{\prime}<4$ and $D_z\geq 0$. As we show below, for  $D_{z}\geq0$ and $h_z=0$ the ground state is a helix with a turn angle given by $\cos \phi = -j_z/4$, where $j_z = J_z/J^{\prime}_z$. For $D_z > 0$, we expect the helix to align and distort to accommodate the easy-axis anisotropy [gray region of the phase diagram in Fig.~3(a)], and we do not have an analytical expression for this state.  Nevertheless, as discussed in the main text, for $D_z$ greater than a lower bound, $D_z^{\text{lb}}$, the ground state is FM for $j_z \leq -2$, $\bm{\tau}_{\sfrac{1}{2}}$-type AF with $\bm{\tau}_{\sfrac{1}{2}}=(0,0,\frac{1}{2})$ for $-2\leq j_z \leq 2$, and A-type AF [$\bm{\tau}_{\text{A}}=(0,0,1)$] for $j_z\geq2$.

We now consider $h_z>0$ and the following variational state [see Fig.~\ref{FigS1}(b)]:
\begin{equation}
{\mathbf S}=(\cos\theta\cos[n\phi],\cos\theta\sin[n\phi],\sin\theta)\,,
\label{eq:S1}
\end{equation}
where $\phi$ is the turn angle, $n$ is an integer representing the layer number along $\mathbf{c}$, and $\theta$ is the canting angle (measured from the $\mathbf{ab}$ plane).  We expect Eq.~(\ref{eq:S1}) to describe the ground state for $h_z \gg S D_z$.

The energy of a state given by Eq.~(\ref{eq:S1}) is 
\begin{eqnarray}
E_2=&&NS^2[J_z(\cos^2\theta\cos\phi+\sin^2\theta)- D_z\sin^2\theta\nonumber\\
+&&J_z^{\prime}(\cos^2\theta\cos2\phi+\sin^2\theta)]-NSh_z\sin\theta\,,\label{eq:E2}
\end{eqnarray}
which is independent of $n$. To obtain a solution for the turn angle, we minimize Eq.~(\ref{eq:E2}) with respect to $\phi$ which gives
\begin{equation}
-J_z\cos^2\theta\sin\phi-2J_z^{\prime}\cos^2\theta\sin2\phi=0\,.\label{eq:S16}
\end{equation}
Since $\cos\theta\neq 0$ in the canted state, it is safe to cancel the term. Equation~(\ref{eq:S16}) then has the following solution:
\begin{equation}
\phi=\cos^{-1}{\bigg[-\frac{J_z}{4J_z^{\prime}}\bigg]}\,.\label{eq:S17}
\end{equation}
$\phi=0$ ($\phi=\pi$) corresponds to FM-aligned (A-type AF-aligned) layers, and $\phi=\frac{\pi}{2}$ corresponds to $\bm{\tau}_{\sfrac{1}{2}}$-type AF-aligned layers.  Other values of $\phi$ correspond to a helix state, or a single-$\mathbf{Q}$ helix state for the case of $D_{z}$ and $D_{xy}=0$.  Note that $\phi$ is independent of $D_z$ and $h_z$.

Next, to determine $h_{\text{sf}}$ and $h_{\text{sat}}$, we minimize Eq.~(\ref{eq:E2}) with respect to $\theta$ and use Eq.~(\ref{eq:S17}) to substitute for $\phi$ in subsequent calculations. From Eq.~(\ref{eq:E2}) we find:
\begin{align}
\theta_1&=\pi/2 \intertext{and}
\theta_2&=\sin^{-1}{\bigg[\frac{4h_zJ_z^{\prime}}{-8SD_zJ_z^{\prime}+S(J_z+4J_z^{\prime})^2}\bigg]}\,.\label{eq:E2_theta_cant}
\end{align}
The first solution is the saturated state, and the second corresponds to a canted-helix state. 

To determine $h_{\text{sf}}$, we need to compare the energy of the canted-helix state to the $h_z=0$ ground-state energy. We assume that the orientation of the spins in the spin-flopped state is given by Eq.~(\ref{eq:S1}). The expressions for $h_{\text{sf}}$ we derive below are therefore only lower bounds. Since the $h_{z} =0$ ground state with $D_z \gg J_z$ is A-type AF [$\bm{\tau}_{\text{A}}=(0,0,1)$] for  $2 < j_z < 4$ and $\bm{\tau}_{\sfrac{1}{2}}$-type AF [$\bm{\tau}_{\sfrac{1}{2}}=(0,0,\frac{1}{2})$] for $-2< j_z < 2$, we consider the two cases separately. 

\subsubsection{$(0,0,1)$ order for $2<J_z/J_z^{\prime}<4$, $D_z \geq 0$, and $h_z>0$}
Substituting $\phi=\pi$ and $\theta=0$ into Eq.~(\ref{eq:E2}) and taking $h_z=0$ yields a zero-field ground state energy for the A-type AF order of
\begin{equation}
E=NS^2(-J_z-D_z+J_z^{\prime})\,.\label{eq:E2_Atype}
\end{equation}
Upon setting Eqs.~(\ref{eq:E2_Atype}) and (\ref{eq:E2}) equal to each other, and substituting Eq.~(\ref{eq:S17}) for $\phi$ and Eq.~(\ref{eq:E2_theta_cant}) for $\theta$, we find that
\begin{equation}
h_{\text{sf}}=\frac{SJ_{z}^{\prime}}{4}\sqrt{[8d_{z}- (j_{z}-4)^2][-8d_{z}+ (j_{z}+4)^2]} \,,\label{eq:S21}
\end{equation}
where $j_{z} = J_{z}/J_{z}^{\prime}$ and $d_{z} = D_{z}/J_{z}^{\prime}$. Similarly to the previous section, this spin-flop field is only valid if $\sin\theta<1$, otherwise the system directly saturates. This happens when
\begin{equation}
h_{\text{sat}}=2SJ_z\,.\label{eq:S22}
\end{equation}
Upon setting Eq.~(\ref{eq:S22}) equal to the value of field that solves Eq.~(\ref{eq:E2_theta_cant}) for $\theta=\frac{\pi}{2}$, we arrive at the condition for the transition:
\begin{equation}
j_z^2=8\left(d_z-2\right)\,.
\end{equation}
To summarize, in the $2<J_z/J_z^{\prime}<4$ and $D_{z}>0$ part of the phase diagram we have the following spin-flop and saturation fields:
\begin{align}
&h_{\text{sat}}=2SJ_z  \intertext{for $j_z^2< 8\left(d_z-2\right)$, and}
&h_{\text{sf}}=\frac{SJ_{z}^{\prime}}{4}\sqrt{8d_{z}- (j_{z}-4)^2} \sqrt{-8d_{z}+ (j_{z}+4)^2}\,,\\ &h_{\text{sat}}=\frac{SJ_{z}^{\prime}}{4}\left[-8d_{z}+(j_{z}+4)^2\right]\,,
\end{align}
for $j_z^2>8\left(d_z-2\right)$.

\subsubsection{$(0,0,\frac{1}{2})$ order for $-2<J_z/J_z^{\prime}<2$}
Substituting $\phi=\frac{\pi}{2}$ and $\theta=0$ into Eq.~(\ref{eq:E2}) and taking $h_z=0$ yields a zero-field ground state energy for the $\bm{\tau}_{\sfrac{1}{2}}$-type AF order of
\begin{equation}
E=NS^2(-D_z-J_z^{\prime})\,.\label{eq:E2_half}
\end{equation}
Setting Eq.~(\ref{eq:E2_half}) equal to Eq.~(\ref{eq:E2}), and substituting Eq.(~\ref{eq:S17}) for $\phi$ and Eq.~(\ref{eq:E2_theta_cant}) for $\theta$ gives
\begin{equation}
h_{\text{sf}} =\frac{S J_{z}^{\prime}}{4}\sqrt{8d_{z}-j_{z}^2} \sqrt{-8d_{z}+(j_{z}+4)^2}\,. \label{eq:S28}
\end{equation}

This spin-flop field is only valid if $\sin\theta<1$, otherwise the system directly saturates. This happens when:
\begin{equation}
h_{\text{sat}}=S(J_z+2J_z^{\prime})\,.\label{eq:S29}
\end{equation}
Upon setting Eq.~(\ref{eq:S29}) equal to the value of field that solves Eq.~(\ref{eq:E2_theta_cant}) for $\theta=\frac{\pi}{2}$, we arrive at the condition for the transition:
\begin{equation}
j_z=-2\pm2\sqrt{2d_z-1}\,.
\end{equation}
Since this calculation is only valid for $j_z>-2$, we disregard the solution with the minus sign.

To summarize, in the $-2<j_z<2$ and $D_{z}>0$ part of the phase diagram we have the following spin flop and saturation fields:
\begin{align}
&h_{\text{sat}} =S(J_z+2J_z^{\prime}) \label{eq:S2} \intertext{for $j_z<-2+2\sqrt{2d_z-1}$, and}
&h_{\text{sf}}  =\frac{S J_{z}^{\prime}}{4}\sqrt{8d_{z}-j_{z}^2} \sqrt{-8d_{z}+(j_{z}+4)^2}\,, \label{eq:S3}\\
&h_{\text{sat}} =\frac{SJ_{z}^{\prime}}{4}\left[-8d_{z}+(j_{z}+4)^2\right] \label{eq:S4}
\end{align}
for $j_z>-2+2\sqrt{2d_z-1}$.  Equations~(\ref{eq:S3}) and (\ref{eq:S4}) also apply for $d_z < 1/2$. 

\subsubsection{Comparison with experimental data}
Using the expressions for $2 <J_{z}/J_{z}^{\prime} < 4$, the experimental $M(H)$ data for $x=0$~\cite{Sangeetha_2017} and experimentally determined value of $\mu=0.43~\mu_{\text{B}}/\text{Co}$~\cite{Jayasekara_2017}, we estimate that $J_{z} \approx 0.24$~meV and $D_{z} \approx 0.08$~meV for $x=0$.  We place $x=0$ on the phase diagram in Fig.~3(a) by arbitrarily assuming that $J_{z}/J_{z}^{\prime}=3$.  

We place the $x=0.24$ and $x=0.44$ compounds in the $-2<J_{z}/J_{z}^{\prime}<2$ region of the phase diagram in Fig.~3(a) based on our neutron diffraction result that both compositions have $\bm{\tau}=\bm{\tau}_{\sfrac{1}{2}}$, with $\bm{\mu}\parallel\mathbf{c}$ for $x=0.24$ (corresponding to $D_z > 0$) and $\bm{\mu}\perp\mathbf{c}$ for $x = 0.44$ (corresponding to $D_z < 0$). Further, we place the $x=0.24$ compound on the FM side ($J_z < 0$) and close to the FM boundary ($J_z/J^{\prime}_z \gtrsim -2$), as $M(H)$ curves for $x=0.25$ are quite soft, showing small saturation fields~\cite{Sangeetha_2017}. We think that the proximity to the $j_z = -2$ phase boundary and/or the frustrated FM NL exchange combined with sufficiently large $d_z$ may explain the reported strong FM correlations along $\mathbf{c}$ in the midst of AF order, as well as a value for $\mu$ lower than that found for either $x=0$ or $0.44$.  Magnetization data for $x=0.45$ do not show evidence for strong FM correlations along $\mathbf{c}$~\cite{Sangeetha_2017}, hence we place $x=0.44$ on the positive side of the $J_{z}/J^{\prime}_{z}$ axis.

\subsection{Estimation of  $\bm{D_{xy}^{\text{lb}}}$}
In the case of easy-plane anisotropy corresponding to spins lying in the $\mathbf{ab}$ plane ($D_z < 0$), the ground state for $D_{xy} = 0$ is a helix with $\bm{\mu}$ lying in the $\mathbf{ab}$ plane. In order to obtain the experimentally observed $\bm{\tau}$ of $\bm{\tau}_{\sfrac{1}{2}} = (0,0,\frac{1}{2})$ for $x=0.44$, $D_{xy}$ must be finite and larger than a lower bound of $D_{xy}^{\text{lb}}$.  To find $D_{xy}^{\text{lb}}$, two calculations are necessary: ($1$) we need to determine the energy difference between the helix state and the $\bm{\tau}_{\sfrac{1}{2}}$-type AF state (i.e.\,the energy gap $\delta E_{\text{gap}}$ to overcome); ($2$) we need to determine how $D_{xy}$ affects the energy of the helix state versus how it affects the $\bm{\tau}_{\sfrac{1}{2}}$-type AF state.  Namely, we need to determined how effective $D_{xy}$ is at overcoming $\delta E_{\text{gap}}$.

For ($1$), we use the above results to determine the energy of the FM, $\bm{\tau}_{\sfrac{1}{2}}$-type AF, A-type AF, and helix states to be, respectively:
\begin{align}
&E_0=NS^2(J_z + J_z^{\prime})\,,\\
&E_\frac{\pi}{2}=-NS^2J_z^{\prime}\,,\\
&E_\pi=NS^2(-J_z + J_z^{\prime})\,,\intertext{and}
&E_{\text{helix}}=NS^2(J_z\cos{\phi} + J_z^{\prime} \cos{2\phi})\,.
\end{align}
As in Eq.~(\ref{eq:S17}), $\cos{\phi}=-\frac{J_z}{4J_z^{\prime}}$ gives the turn angle for the helix. Whether the FM, $\bm{\tau}_{\sfrac{1}{2}}$-type AF, or A-type AF state is closest in energy to the helix state is dependent on the $J_z/J_z^{\prime}$ ratio.  The gaps are:
\begin{align}
&\delta E_{\text{gap}}=E_0-E_{\text{helix}} \text{ for } J_z/J_z^{\prime}<-2\,,\label{eq:EgapE0}\\ 
&\delta E_{\text{gap}}=E_\frac{\pi}{2}-E_{\text{helix}} \label{eq:EgapEpiov2} \text{ for } -2<J_z/J_z^{\prime}<2 \intertext{and}
&\delta E_{\text{gap}}=E_\pi-E_{\text{helix}} \text{ for } J_z/J_z^{\prime}>2\,. \label{eq:EgapEpi}
\end{align}

For ($2$), we look at the energy contribution of $D_{xy}$ in terms of unit strength. For the FM, $\bm{\tau}_{\sfrac{1}{2}}$-type AF, or A-type AF state, we get, by design, energies of
\begin{equation}
E_{0,\frac{\pi}{2},{\pi}}^{D_{xy}}=E_{0}^{D_{xy}}=E_{\frac{\pi}{2}}^{D_{xy}}=E_{\pi}^{D_{xy}}=-1\,,
\end{equation}
where we have dropped the factor of $NS^2$ for convenience.  For the helix state we need to calculate:
\begin{equation}
E_{\text{helix}}^{D_{xy}}=-\frac{1}{N}\sum_{n=0}^{N-1} \sin^4{n\phi}+\cos^4{n\phi}\,,
\end{equation}
where we sum over $N$ layers that make a full turn of the helix commensurate with the chain. The summation can be done analytically by using the trigonometric identities
\begin{align}
&\sin^4{n\phi}=\frac{1}{8}(3-4\cos{2n\phi}+\cos{4n\phi})\intertext{and}
&\cos^4{n\phi}=\frac{1}{8}(3+4\cos{2n\phi}+\cos{4n\phi})\,,
\end{align} 
and noting that the sum over the linear trigonometric functions are averages over the period, which is zero. Thus, we only have the constant terms left, and $E_{\text{helix}}^{D_{xy}}=-\frac{3}{4}$. Therefore, the unit strength $D_{xy}$ term creates an energy difference of $\delta E_{D_{xy}}=E_{\text{helix}}^{D_{xy}}-E_{0,\frac{\pi}{2},{\pi}}^{D_{xy}}=\frac{1}{4}$, and
\begin{equation}
D_{xy}^{lb}=\frac{\delta E_{\text{gap}}}{\delta E_{D_{xy}}}=4\delta E_{\text{gap}}\,,
\end{equation}
where $\delta E_{\text{gap}}$ is given in Eqs.~(\ref{eq:EgapE0})--(\ref{eq:EgapEpi}). $D^{lb}_{xy}(J_z/J^{\prime}_z)$ is shown in Fig.~3(c).

\subsection{Estimation of $\bm{D_{z}^{\text{lb}}}$}

The calculation of the equivalent lower bound on $D_z$ follows a similar vein. The only differences are due to the effect of the unit strength of $D_z$ on the helix state, and that we assume that the plane of the helix contains $\mathbf{c}$.  The energy we need to calculate to determine the unit strength of $D_{z}$ is 
\begin{eqnarray}
E_{\text{helix}}^{D_z}&&=-\frac{1}{N}\sum_{n=0}^{N-1} \cos^2{n\phi}=-\frac{1}{N}\sum_{n=0}^{N-1} \frac{1}{2}(1+\cos{2n\phi})\nonumber\\&&=\frac{1}{2},
\end{eqnarray}
which leads to
\begin{equation}
D_z^{\text{lb}}=\frac{\delta E_{\text{gap}}}{\delta E_{D_z}}=2\delta E_{\text{gap}}\,.
\end{equation}
Thus, the shape of the $D_{z}^{\text{lb}}(J_{z}/J^{\prime}_{z})$ curve is the same as $D_{xy}^{\text{lb}}(J_{z}/J^{\prime}_{z})$, but has half the magnitude.  $D_{z}^{\text{lb}}(J_{z}/J^{\prime}_{z})$ is plotted in Fig.~3(c).

\bibliographystyle{apsrev4-1.bst}
\bibliography{Sr_doped_CaCo2As2_diffraction.bib}

\end{document}